\documentclass{iopart}

\usepackage{ifthen}
\usepackage{ifpdf}

\usepackage{latexsym}
\usepackage{amssymb} 
\usepackage{bm}

\ifpdf
\usepackage{graphicx}
\usepackage{epstopdf}
\else
\usepackage{graphicx}
\usepackage{epsfig}
\fi

\newcommand{\const}{\mbox{const}}

\newcommand{\tbox}[1]{\mbox{\tiny #1}}

\newcommand{\be}[1]{\begin{eqnarray}\ifthenelse{#1=-1}{\nonumber}{\ifthenelse{#1=0}{}{\label{e#1}}}}
\newcommand{\ee}{\end{eqnarray}} 

\newcommand{\DE}{D} 
\newcommand{\sect}[1]{\section{#1}}
\newcommand{\hide}[1]{\textcolor{red}{[hidden text]}}

\begin{document} 

\title[Anomaly]{Quantum anomalies and linear response theory} 

\author{\mbox{Itamar Sela$^{1}$, James Aisenberg$^{2}$, Tsampikos Kottos$^{2}$, Doron Cohen$^{1}$}}

\address{
$^1$Department of Physics, Ben-Gurion University, Beer-Sheva 84105, Israel\\
$^2$Department of Physics, Wesleyan University, Middletown, CT 06459, USA}

\begin{abstract}
The analysis of diffusive energy spreading  
in quantized chaotic driven systems,  
leads to a universal paradigm for the emergence 
of a quantum anomaly.
In the classical approximation a driven chaotic system 
exhibits stochastic-like diffusion in energy space 
with a coefficient~$D$ that is proportional 
to the intensity $\varepsilon^2$ of the driving.
In the corresponding quantized problem the 
coherent transitions are characterized by 
a generalized Wigner time $t_{\varepsilon}$, 
and a self-generated (intrinsic) dephasing process 
leads to non-linear dependence of $D$ on $\varepsilon^2$.
\end{abstract}


A major theme in mechanics concerns the response 
of a system to a driving source $f(t)$,  
given that the interaction term is $\mathcal{H}_{\tbox{int}}=f(t)V$.
This leads to the well known framework 
of linear response theory (LRT) with  
its celebrated fluctuation-dissipation relation.   
Below we assume a stationary driving source which is characterized by 
a power spectrum $\tilde{S}(\omega)=\mbox{FT}[\langle f(t)f(0) \rangle]$, 
where FT stands for Fourier transform.
In the absence of driving the stationary fluctuations of the system 
are characterized by the spectral function $\tilde{C}(\omega)=\mbox{FT}[\langle V(t)V(0) \rangle]$.
In the presence of driving the main three effects are:  
the {\em decay} of the initial preparation; 
the spreading and eventually the {\em diffusion} in energy space;
and the associated {\em heating}.

\sect{LRT and Kubo} 
Strict LRT behavior means that the diffusion in energy space \cite{ott} 
and the related absorption coefficient \cite{wilk,jar,frc} 
are {\em linear functional}  of the spectral function $\tilde{S}(\omega)$.  
Specifically, the Kubo formula for the diffusion coefficient in energy space is   
\be{1}
\DE \ \ = \ \ \frac{1}{2}\int_{-\infty}^{\infty} 
\omega^2 d\omega \ \tilde{C}(\omega) \tilde{S}(\omega)
\ee
It follows that the diffusion is proportional 
to the intensity of the driving~$\varepsilon^2$ 
as defined below. 
We are going to consider on equal footing  
driving by a quasi-constant perturbation $f(t)\,{\sim}\const$,  
and quasi-linear DC driving with $\dot{f}(t)\,{\sim}\const$.
The notation ``${\sim}\const$'' means that it is constant  
over large time intervals of duration~$t_{\varphi}$, 
with some characteristic RMS value that we call $\varepsilon$. 
Accordingly the associated spectral functions is 
\be{2}
\tilde{S}(\omega) \ \ = \ \ \varepsilon^2 \omega^{-\sigma} \delta_{\gamma}(\omega)
\ee
where $\delta_{\gamma}(\omega)=(\gamma/\pi)/(\omega^2+\gamma^2)$ 
with $\gamma=1/t_{\varphi}$, 
while the spectral exponent is $\sigma{=}0$ for quasi-constant perturbation, 
and  $\sigma{=}2$ for quasi linear DC driving.

\sect{Paradigms} 
We are looking for circumstances where the Kubo formula  
is not applicable. This means that either $\DE$ 
is still proportional to the strength of the driving 
but with an anomalous coefficient, or more generally $\DE$ 
might depend in a non-linear way on the strength of the driving.  
Paradigms for non-LRT response with which we are familiar are:   
{\bf \ (i)} Classical non-LRT route that follows 
from the Kolmogorov-Arnold-Moser scenario as in the 
driven (kicked) rotator problem \cite{ll};  
{\bf \ (ii)} Quantum semi-LRT due to sparsity and textures 
that characterize the perturbation matrix \cite{slr};
{\bf \ (iii)} Quantum corrections due to dynamical 
localization effect \cite{fishman,kravtzov}.
{\bf \ (iv)} Quantum absorption due to Landau-Zener transitions 
between neighboring energy levels \cite{wilk}; 
{\bf \ (v)} Quantum non-perturbative anomalies 
that are associated with having a finite spectral bandwidth \cite{rsp}.

\sect{Scope and main observation} 
One should notice that all the mentioned non-LRT paradigms above 
become irrelevant if we consider the {\em continuum} limit 
of a {\em universal} quantized chaotic system. 
By definition ``continuum'' means that 
the Heisenberg time (see definition later) can be 
taken as infinite, while ``universality'' means 
that the correlation time (see definition later)   
can be taken as zero. It should be clear that our considerations 
apply to strictly chaotic systems that do not have mixed phase-space dynamics.  
In the present work we show that even with all these 
assumptions and exclusions there is still room for 
a novel manifestation of quantum mechanical anomalies 
in the response to external driving.  
A central role is played by the generalized Wigner time  $t_{\varepsilon}$
which characterizes a coherent spreading process in energy space, 
and by what we call intrinsic dephasing time $t_{\varphi}^{\tbox{(eff)}}$.

Beyond any technical details it is important to realize that for a universal chaotic
system, as assumed in Random Matrix Theory (RMT) studies, 
the Kubo formula of LRT can be deduced merely via {\em dimensional analysis}.
The non-Ohmic generalization of this statement [Eq.(\ref{e19})], 
as established in this communication, implies a universal quantum anomaly
in the response characteristics of non-Ohmic systems.
This prediction has potential applications e.g. with regard to the rate of heating
of cold atoms in vibrating traps.

\sect{Modeling} 
In a system that is described by a time dependent 
Hamiltonian $\mathcal{H}[R]$ with $R=R_0{+}f(t)$ 
the transitions between the adiabatic energy 
levels $E_n$ are induced by the perturbation 
matrix $V_{nm}=(d\mathcal{H}/dR)_{nm}$.   
The spectral function that characterizes the fluctuations 
of~$V$ in the absence of driving is
\be{3}
\tilde{C}(\omega)  = \sum_n |V_{n,n_0}|^2 2\pi\delta\left(\omega -\frac{E_n{-}E_{n_0}}{\hbar}\right)  
\ee
with implicit average over $n_0$ as determined 
by the energy window of interest. We assume below that  
\be{6}
\tilde{C}(\omega) = 2\pi|\omega|^{s_0-1} 
\ \ \ \ \ \ \ \ \ \mbox{for $\omega_0  < |\omega| < \omega_{\tbox{cl}}$}
\ee
and distinguish between the Ohmic ($s_0{=}1$), subOhmic ($0{<}s_0{<}1$), 
and superOhmic ($1{<}s_0{<}2$) cases.    
Without loss of generality, by appropriate rescaling of~$f(t)$, 
we set the prefactor in Eq.(\ref{e6}) as $2\pi$. 
The infrared cutoff ${\omega_0 =(\hbar\varrho)^{-1}}$ 
is the mean level spacing, as determined by the 
density of states. The ultraviolet 
cutoff $\omega_{\tbox{cl}}$ is determined by 
the classical dynamics and is 
known as the bandwidth or as the 
ballistic version of the Thouless energy. 
The associated time scales are the 
Heisenberg time $t_{\tbox{H}} = 2\pi\hbar\varrho$ 
and the classical correlation time $t_{\tbox{cl}} = 2\pi/\omega_{\tbox{cl}}$. 
In a later paragraph we define the generalized Wigner 
time $t_{\varepsilon}$  that depends on the strength of the driving.
This time scale characterizes the coherent spreading process. 
We assume below mesoscopic circumstances for which 
\be{0}
t_{\tbox{cl}} \ \ll \ (t_{\varepsilon}, t_{\varphi}) \ \ll \ t_{\tbox{H}}
\ \ \ \ \ \ \ \ \mbox{[mesoscopics]}
\ee  
where $t_{\varphi}$ is the correlation time of the 
driving source as define with relation to Eq.(\ref{e2}).
Our interest is in results that remain well defined 
for ${t_{\tbox{cl}}\rightarrow0}$ (universal limit) 
and ${t_{\tbox{H}}\rightarrow\infty}$ (continuum limit). 
The existence of a universal limit is the underlying 
postulate in RMT modeling.  
We distinguish in the analysis between 
weak (${t_{\varepsilon}>t_{\varphi}}$) 
and strong (${t_{\varepsilon}<t_{\varphi}}$) driving.

\sect{DC driving} 
The common interest is in Ohmic systems (${s_0=1}$) 
with quasi-linear DC driving (${\sigma=2}$), 
for which Kubo formula gives $\DE=\pi\varepsilon^2$. 
This result is {\em not} sensitive to $t_{\varphi}$,
and is independent of the infrared and ultraviolet 
cutoffs. Our purpose is to generalize this result 
for the case of non-Ohmic fluctuations. 
We shall see that this requires to go beyond LRT.

The key observation is that the problem of quasi-linear DC driving
reduces, with some reservations, to the analysis of 
quasi-constant perturbation. This is done by transforming 
the Hamiltonian into the adiabatic basis where it takes the form  
\be{9}
\tilde{\mathcal{H}} = \mbox{diag}\{E_{n}\} 
+ \dot{f} \left\{i\frac{\hbar V_{nm}}{E_{n}-E_{m}} \right\} 
\ee
If we ignore the implicit time dependence of the adiabatic  
energies and matrix elements, then this Hamiltonian is the same 
as that of quasi-constant perturbation but with 
{\em effective} exponent ${s=s_0-2}$. 
In particular quasi-linear driving of 
an Ohmic system corresponds to ${s=-1}$.

At this point one wonders what is the effect of the 
residual implicit time dependence of the Hamiltonian Eq.(\ref{e9}). 
Obviously we should not be too worried about the wiggles 
of the levels, because they take place on a very small 
energy scale and would be of relevance only if we were 
considering times of the order of the Heisenberg time.
On the other hand, the variation of the matrix elements cannot be ignored,
and we shall come back to this issue later.

\sect{The generalized Wigner time} 
Universality (irrelevance of $\omega_{\tbox{cl}}$) 
is a common built-in assumption in numerous ``quantum chaos" studies 
that utilize the standard random matrix ensembles. 
Furthermore a quasi-continuum assumption (irrelevance of  $\omega_0$) 
is implicit in the standard derivations of LRT. 
If we believe that in the {\em continuum} limit ($\omega_0\rightarrow0$) 
there exists a {\em universal} limit ($\omega_{\tbox{cl}}\rightarrow\infty$) 
that leads to a generalized response theory, 
then disregarding $t_{\varphi}$  the only relevant time scale    
that might emerge in the dynamics is implied by dimensional analysis: 
\be{7}
t_{\varepsilon} \ \ = ({\hbar}/{\epsilon})^{2/(2{-}s)} 
\ee  
One immediately realizes that it is the generalized Wigner time 
of Ref.\cite{kbs}. For a quasi-constant perturbation 
of Ohmic system it is literally the Wigner time 
\be{10}
t_{\varepsilon}=(\hbar/\Gamma_{\tbox{E}}), 
\ \ \ \ \ \ \ \ \mbox{for ${s=1}$}
\ee  
where ${\Gamma_{\tbox{E}}=(2\pi/\hbar)\epsilon^2}$ 
is the Fermi-golden-rule rate of transitions.  
For a quasi-linear driving of Ohmic system it is
the breaktime scale that has been 
introduced in Ref.\cite{frc},  
\be{11}
t_{\varepsilon}=(\hbar^2/\DE)^{1/3}, 
\ \ \ \ \ \ \ \ \mbox{for ${s=-1}$}
\ee  
where $\DE=\pi\varepsilon^2$ in derived from the Kubo formula.

More generally, in the non-Ohmic case, we can associate 
with the generalized Wigner time an energy scale~$\hbar/t_{\varepsilon}$ 
and a diffusion coefficient  
\be{19}
\DE_{\varepsilon} \ = \ 
\frac{\hbar^2}{t_{\varepsilon}^3}
\ = \ 
\hbar^{-2\frac{s{+}1}{2{-}s}}
\ \varepsilon^{\frac{6}{2{-}s}},
\ \ \ \ \ \ \ \ \mbox{[${s=s_0{-}\sigma}$]}
\ee  
At this point one should wonder whether this 
expression might emerge from the analysis 
of the spreading process in some universal limit.
Note that it is only in the Ohmic (Kubo) case 
that $\DE_{\varepsilon}$ becomes $\hbar$ independent.

\sect{Coherent spreading}   
If the perturbation matrix in Eq.(\ref{e9}),  
call it $W_{nm}$, were strictly time independent, 
then the induced wavepacket dynamics would lead 
to a steady state, with a saturation profile that 
reflects the local density of states. 
Specifically, let us assume that the system is prepared in 
the unperturbed state $n$ for which the unperturbed energy $E_{n}$ 
is well defined, then in the perturbed basis 
it has the energy distribution 
\be{0}
P_{\infty}(E)= \sum_{\nu} |\langle \nu|n\rangle|^2\delta(E{-}E_{\nu})
\ee
We have argued in Ref.\cite{kbs}, following previous studies,  
that this energy distribution has a semicircle-like 
core that extends within $|E{-}E_n|<\hbar/t_{\varepsilon}$, 
coexisting with outer perturbative tails 
that are determined by the first order 
expression $|W_{nm}|^2/(E_n{-}E_m)^2$
for the overlaps. The associated variance 
is $\Delta E(\infty)^2 = \omega_{\tbox{cl}}^{s} \varepsilon^2$ for $s>0$, 
and $\Delta E(\infty)^2 = (\hbar/t_{\varepsilon})^2$ for $s<0$.

In the time dependent analysis the steady 
state profile of $P_t(E)$  is achieved only 
after $t_{\varepsilon}$, but the crossover 
is not necessarily observed in the spreading $\Delta E(t)$, 
which is a second moment calculation. 
Specifically we get 
\be{15}
\Delta E(t) =
& \varepsilon \ \omega_{\tbox{cl}}^{s/2} \ \ \ \ \ \ 
& \mbox{for [$s{>}0$],[$t{>}t_{\tbox{cl}}$]}
\\
\label{e15b}
\Delta E(t) =
& \varepsilon \ t^{|s|/2}  \ \ \ \ \ \  
& \mbox{for [$s{<}0$],[$t_{\tbox{cl}}{<}t{<}t_{\varepsilon}$]}
\\
\label{e15c}
\Delta E(t) =
& \hbar/t_{\varepsilon} \ \ \ \ \ \ 
& \mbox{for [$s{<}0$],[$t{>}t_{\varepsilon}$]}
\ee  
For ${s>0}$ the spreading saturates 
as soon as $t>t_{\tbox{cl}}$, and to detect the  
crossover at $t_{\varepsilon}$ one should look 
on the survival probability or on percentiles of 
the distributions as described in Ref.\cite{kbs}.  
But for ${s<0}$ the second moment of the evolving 
distribution exhibits the crossover to saturation 
at $t_{\varepsilon}$, and not at $t_{\tbox{cl}}$. 
This can be deduced using the following simple reasoning. 
First order perturbation theory implies 
that the tail grows like $|W_{nm}t|^2$ within 
the shrinking interval ${\hbar\gamma(t)<|E-E_n|<\hbar/t}$.
In the outer ${|E{-}E_n|>\hbar/t}$ region 
the tail is saturated due to recurrences.
The lower cutoff $\gamma(t)$ is determined by a self-consistency condition, 
saying that the integral over the 1st order tail 
of $P_t(E)$ from  $\gamma(t)$ to infinity should be $\mathcal{O}(1)$. 
This leads to the estimate ${\gamma(t)=\left[t^{|s|} + (\varepsilon t/\hbar)^{-2} \right]^{-1/|s|} }$.  
Steady state is achieved at $t_{\varepsilon}$ 
when ${\gamma(t) \sim 1/t}$.  
Up to this time the second moment 
of the evolving distribution is dominated by 
the growing piece of the tail leading to Eq.(\ref{e15b}). 
This is a diffusive-like growth (${\Delta E(t) \propto t^{1/2}}$) 
in the case of a quasi-linear DC driving.

\sect{Diffusion} 
The dephasing time $t_{\varphi}$ indicates the crossover 
from coherent to stochastic spreading behaviour. 
The central limit theorem implies that the long time 
spreading is diffusive with coefficient  
\be{16}
\DE = \frac{\Delta E(t_{\varphi})^2}{2t_{\varphi}} 
\ee  
where $\Delta E(t)$ is the time dependent coherent spreading 
that we have discussed in the previous paragraph.
This result is classical in nature (no $\hbar$) and 
it is easily checked that it agrees with the Kubo formula Eq.(\ref{e1}) 
provided we use Eq.(\ref{e15}) or Eq.(\ref{e15b}), leading to  
\be{17}
\DE \ \ =
& \varepsilon^2 \ \omega_{\tbox{cl}}^{s}  \ t_{\varphi}^{-1} \ \ \ \ \ \ \ \ \ 
&\mbox{for [$s{>}0$]}
\\
\label{e17b}
\DE \ \ =
& \varepsilon^2 \ t_{\varphi}^{|s|{-}1} \ \ \ \ \ \ \ \ \
&\mbox{for [$s{<}0$],[$t_{\varphi}{<}t_{\varepsilon}$]}
\ee  
Both results are $\propto \varepsilon^2$. 
In particular Eq.(\ref{e17}) applies to 
quasi-constant perturbation and is merely 
the well known {\em hopping} estimate for 
the noise-induced diffusion is system 
with `localization'.

Consider a general system with quasi-linear DC driving,  
such that ${s<0}$.
If the driving is weak (${t_{\varphi} < t_{\varepsilon}}$)
it is justified to substitute Eq.(\ref{e15b}) in Eq.(\ref{e16}), 
thus getting the LRT result Eq.(\ref{e17b}). 
In particular we note that in the standard case 
of DC-driven Ohmic system (${s=-1}$) 
we get a $t_{\varphi}$ independent result.
Otherwise the result is $t_{\varphi}$ dependent. 
For ${|s|\ne1}$ one observes that in the 
limit ${t_{\varphi}\rightarrow \infty}$
the LRT result is either zero or infinity.  
This suggests that in such circumstances 
a realistic theory should lead to an $\hbar$~dependent result.

\sect{Beyond LRT} 
So far the elaborated spreading picture that we have introduced 
gave for $\DE$ the same result as Kubo. 
So now we would like to see whether there are circumstances 
where this picture leads to novel physics beyond LRT.
Considering ${s<0}$ and strong driving ${t_{\varepsilon}<t_{\varphi}}$,
it seems that one should substitute Eq.(\ref{e15c}) in Eq.(\ref{e16}),
leading to a sub-linear dependence on the strength of the driving: 
\be{18}
\DE \ \ = \ \ 
\hbar^{\frac{2|s|}{2{+}|s|}} 
\varepsilon^{\frac{4}{2{+}|s|}} 
\ t_{\varphi}^{-1},
\ \ \ \ \ \ \ \ \ [s{<}0],[t_{\varphi}{>}t_{\varepsilon}]
\ee  
However, one should be critical with regard to this result. 
The saturation of the coherent spreading process 
assumes a time-independent perturbation in Eq.(\ref{e9}). 
This would be the case if ${\sigma=0}$ but not if ${\sigma=2}$.

\sect{Intrinsic dephasing hypothesis} 
Considering the general problem of having a driving source 
with arbitrary spectral exponent~$\sigma$,   
one realizes that as far as the integrand of the Kubo formula Eq.(\ref{e1}) 
is concerned the effective spectral exponent is ${s=s_0-\sigma}$. 
If a breakdown of this {\em spectral equivalence} rule is observed, 
it constitutes a demonstration of a quantum anomaly.
Indeed in the numerical experiment of Fig.~1, 
we contrast the dynamics which is generated 
by a time-dependent ($\sigma{=}-2$) perturbation, 
with that of a frozen ($\sigma{=}0$) perturbation  
that has the same $s$.  We find that only the latter 
shows the coherent saturation of Eq.(\ref{e15c}). 

Consequently we would like to conjecture that 
the implicit time-independence 
of the perturbation in Eq.(\ref{e9}) leads  
to an intrinsic dephasing time~$t_{\varphi}^{\tbox{eff}}$ 
which is finite in the limit ${t_{\varphi}\rightarrow \infty}$. 
We have in our problem only one time scale, 
the generalized Wigner time, and therefore the 
natural speculation would be $t_{\varphi}^{\tbox{eff}} \sim  t_{\varepsilon}$.
If this speculation is true, 
then the replacement ${t_{\varphi} \mapsto t_{\varphi}^{\tbox{eff}}}$ 
in Eq.(\ref{e18}) leads to the universal result Eq.(\ref{e19}). 
For the standard Ohmic case this implies 
that diffusive spreading persists beyond $t_{\varepsilon}$, 
and that LRT-like result ${\propto \varepsilon^2}$ still holds.
For the non-Ohmics case this implies 
that the sub-diffusive or super-diffusive 
coherent spreading turns into normal diffusion with non-linear 
dependence on ${\varepsilon^2}$.
Both expectations are supported by the numerical experiment 
of Fig.1 that we further discuss and analyze below.

\sect{RMT numerics} 
In order to have a model that captures and tests  
the question of spectral equivalence it is 
convenient to use not the standard Wigner model, 
but rather its parametric invariant variation: 
\be{20}
\mathcal{H}_{ij}(t) \ = \ \mathcal{E}_i\delta_{ij} +  \cos[f(t)] \ V^{(1)}_{ij}  +  \sin[f(t)] \ V^{(2)}_{ij}
\ee
where $V^{(1)}_{ij}$ and $V^{(2)}_{ij}$ are two independent realizations 
of a banded matrix, that has a bandprofile $\lambda |\omega|^{s_{\lambda}-1}$, 
in the sense of Eq.(\ref{e3}), but with ${n\mapsto i}$. 
The energies $E_n(t)$ are obtained via  
diagonalization of $\mathcal{H}(t)$ 
and the perturbation matrix $V_{nm}(t)$ should be written 
in the same basis. The bandprofile of $V_{nm}$ 
is related to that of $V_{ij}$ as discussed in \cite{blm}. 
We have set $\varrho{=}1$ and $\lambda{=}1$ and verified 
that  ${s_0 \approx s_{\lambda}}$.   
We consider DC driving $\dot{f}=\const$, 
and accordingly ${\varepsilon^2 = \lambda \dot{f}^2}$. 

In the time dependent adiabatic basis 
the perturbation matrix in the transformed Hamiltonian Eq.(\ref{e9}) 
is $W_{nm}(t)=i\dot{f}V_{nm}/(E_n{-}E_m)$, 
whose bandprofile is characterized by $s=s_0{-}2$.   
This matrix changes with time but it preserves its 
statistical properties. The question is whether 
its implicit time dependence generates an effective dephasing process.  

The numerical experiment is simple. On the one hand we make 
simulations with the time dependent Hamiltonian $\mathcal{H}$.
On the other hand we use a frozen version of Eq.(\ref{e9}) 
which we write in the $ij$ basis as 
\be{21}
\mathcal{H}_{ij}(\mbox{\small frozen}) 
\ = \ \mathcal{E}_i\delta_{ij} +  U_{ij}  +  \dot{f} \ W_{ij}(0)
\ee
The initial state is assumed to be localized at ${i=0}$, 
and an ensemble average over realizations is taken. 
Comparing the simulations (Fig.1) we deduce that there is intrinsic 
dephasing due to the implicit time dependence of the driving in Eq.(\ref{e9}).

In order to figure out what is the intrinsic dephasing time we plot 
in Fig.2 the scaled diffusion $\DE/\DE_{\varepsilon}$ versus 
the strength of the driving. The Ohmic case as conjectured is `boring'.
In contrast to that the sub-Ohmic and the super-Ohmic case 
exhibit departure from the universal expectation for 
large and small $\varepsilon$ respectively, indicating that the 
effective dephasing time becomes shorter than $t_{\varepsilon}$.
We associate this systematic deviation with the infrared 
and ultraviolet cutoffs respectively: otherwise dimensional
analysis implies that such deviation cannot emerge. 
We explain this sensitivity as follows:  The value of $D$ is 
most sensitive to $t_{\varphi}$ in the slow diffusion stage.
In the sub (super) Ohmic case the slow diffusion stage is 
for long (short) times, and accordingly the sensitivity is 
to the lower (upper) cutoff, in-spite of the fact that the spreading 
is dominated by the high (low) frequency transitions.

\sect{Conclusions} 
LRT gives a finite classical-like result for the response 
of a low frequency driven Ohmic system.
But in the case of a sub-Ohmic or super-Ohmic system, 
classical LRT predicts in the same limit either zero or infinite response. 
This is the notch where quantum-mechanics becomes 
most relevant, leading to an anomalous non-linear response.

The analysis highlights the role which is played by 
the generalized Wigner time of \cite{kbs}, which is the only 
relevant time scale in the universal continuum limit, 
and leads to a single-parameter expression Eq.(\ref{e19}) 
for the diffusion in energy space. 

The results might have a direct application concerning 
the heating rate of cold atoms in vibrating traps \cite{atoms}, 
where the experimentalist has control over both 
the shape (hence $\tilde{C}(\omega)$) 
and the power spectrum ($\tilde{S}(\omega)$) of the driving. 
In particular we note that if cold atoms are ``shaken" without deforming the shape of
the billiard, the spectral function that describes the fluctuations becomes sub-Ohmic \cite{wls}.


\ack

This research was supported by a grant from the USA-Israel Binational Science Foundation (BSF).




\vspace*{1mm}

\begin{figure}[h!t]

\includegraphics[width=0.8\hsize,clip]{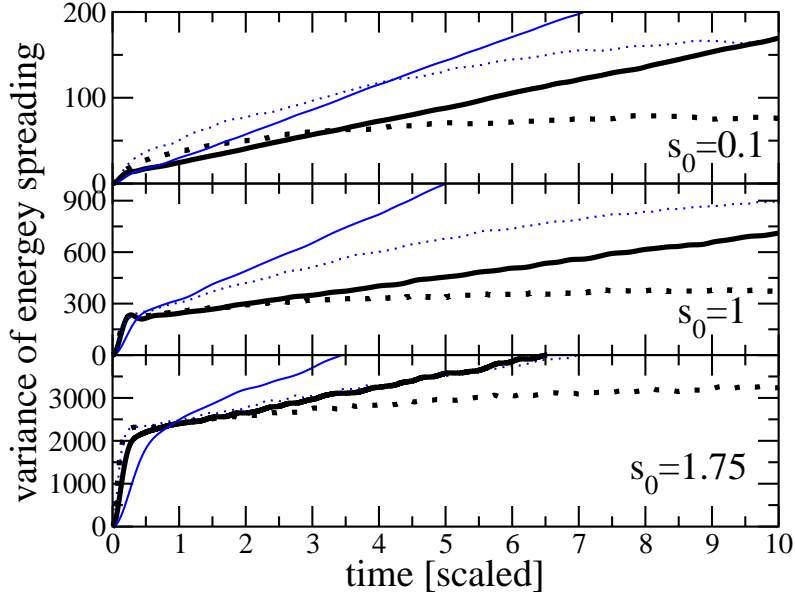}

\caption{(color online)
Variance of energy spreading versus time 
for three different $s_0$-values, 
and two different driving rates 
$\dot{f}{=}5$ (thick black lines) 
and $\dot{f}{=}12$ (thin blue lines). 
The time-axis is rescaled according 
to the Wigner time $t_{\epsilon}$ given by Eq.(\ref{e7}). 
Solid lines correspond to the simulations 
based on the Hamiltonian of Eq.(\ref{e20}), 
while the dotted are for its frozen version Eq.(\ref{e21}). 
} 
\end{figure}

\begin{figure}[h!t]

\includegraphics[width=0.8\hsize,clip]{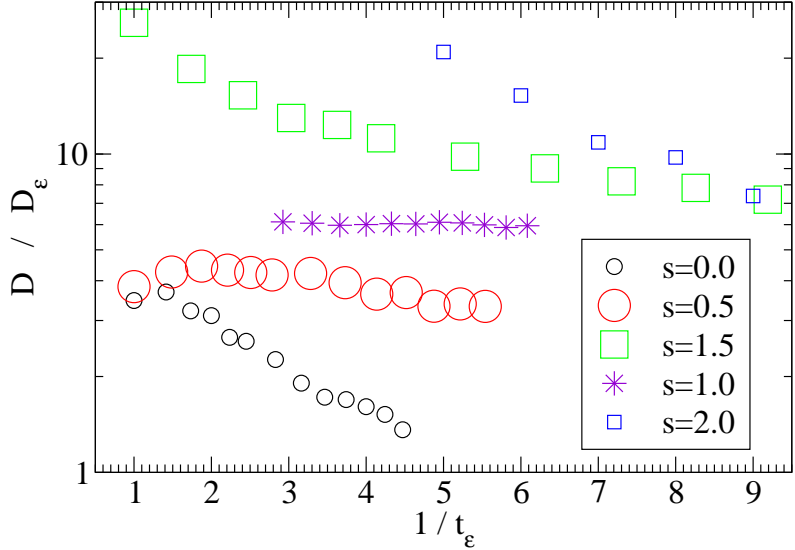}

\caption{(color online)
Dependence of diffusion on the rate of the driving 
for various values of $s_0$.
The axes are $X=\varepsilon^{2/(2{-}s)}$ 
and $Y=D/\varepsilon^{6/(2{-}s)}$, where $s=s_0{-}2$. 
Note that ${\omega_0 < X < \omega_{\tbox{cl}}}$ where 
the level spacing is ${\omega_0=1}$ and the bandwidth is ${\omega_{\tbox{cl}}=50}$. 
The deviation of $D$ from universlity is due 
to the finite infrared or ultraviolet cutoffs: 
We see that in the superOhmic case (${s_0>1}$) 
the diffusion $D$ becomes $\omega_0$ independent 
for large $\varepsilon$, while in the subOhmic case (${s_0<1}$) 
it becomes $\omega_{\tbox{cl}}$ independent for small $\varepsilon$.
} 

\end{figure}

\clearpage

\Bibliography{99}

\bibitem{ott}
E. Ott, Phys. Rev. Lett. {\bf 42}, 1628 (1979). 
R. Brown, E. Ott and C. Grebogi, 
Phys. Rev. Lett, {bf 59}, 1173 (1987); 
J. Stat. Phys. {\bf 49}, 511 (1987). 

\bibitem{wilk}
M. Wilkinson, J. Phys. A {\bf 21}, 4021 (1988); 
J. Phys. A {\bf 20}, 2415 (1987).

\bibitem{jar}
C. Jarzynski, {\em Phys. Rev.} {\bf E 48}, 4340 (1993); 
Phys. Rev. Lett. {\bf 74}, 2937 (1995).

\bibitem{frc} 
D. Cohen, Annals of Physics {\bf 283}, 175 (2000). 

\bibitem{ll} 
A.J. Lichtenberg and M.A. Lieberman, 
{\it Regular and Stochastic Motion}, 
(Springer, Berlin, 1983).

\bibitem{slr}
D. Cohen, T. Kottos and H. Schanz, J. Phys. A {\bf 39}, 11755 (2006).
M. Wilkinson, B. Mehlig and D. Cohen, Europhys. Lett. {\bf 75}, 709 (2006).
See also: A. Stotland, T. Kottos and D. Cohen, Phys. Rev. B {\bf 81}, 115464 (2010).

\bibitem{fishman}
For review and references see lecture notes 
by M. Raizen, in "Quantum Chaos",
{\em Proceedings of the International School
of Physics "Enrico Fermi", Course CXLIII},
Ed. G. Casati, I. Guarneri and U. Smilansky
(IOS Press, Amsterdam 2000).

\bibitem{kravtzov}
D.M. Basko, M.A. Skvortsov and V.E. Kravtsov,
Phys. Rev. Lett. {\bf 90}, 096801 (2003).

\bibitem{rsp} 
D. Cohen, Phys. Rev. Lett. {\bf 82}, 4951 (1999). 
D. Cohen and T. Kottos, Phys. Rev. Lett. {\bf 85}, 4839 (2000).

\bibitem{kbs} 
J. Aisenberg, I. Sela, T. Kottos, D. Cohen and A. Elgart, J. Phys. A {\bf 43}, 095301 (2010); 
I. Sela et al, Phys. Rev. E {\bf 81}, 036219 (2010). 

\bibitem{blm} 
J.A. Mendez-Bermudez, T. Kottos and D. Cohen, 
Phys. Rev. E {\bf 73}, 036204 (2006).

\bibitem{atoms} 
M. Andersen, A. Kaplan, T. Grunzweig, N. Davidson, 
Phys. Rev. Lett. {\bf 97}, 104102 (2006). 
A. Stotland, D. Cohen, N. Davidson, 
Europhys. Lett. {\bf 86}, 10004 (2008). 

\bibitem{wls} 
A. Barnett, D. Cohen and E.J. Heller, 
Phys. Rev. Lett. {\bf 85}, 1412 (2000); 
J. Phys. A {\bf 34}, 413 (2001).

\end{thebibliography}
\end{document}